\documentclass[prl,aps,onecolumn,preprintnumbers,showpacs,superscriptaddress]{revtex4-1}
\usepackage{graphicx}
\usepackage{epsfig}
\usepackage{epsf}
\usepackage{amssymb}
\usepackage{amsmath}
\usepackage{amsthm}
\usepackage{multirow}
\usepackage{hyperref}

\usepackage[usenames]{color}
\definecolor{Red}{rgb}{1,0,0}
\definecolor{Blue}{rgb}{0,0,1}

\begin{document}

\title{Experimental Realization of Quantum Artificial Intelligence}

\author{Zhaokai Li}
\affiliation{Hefei National Laboratory for Physical Sciences at Microscale and Department of Modern Physics, University of Science and Technology of China, Hefei, Anhui 230036, People's Republic of China}
\affiliation{CAS Center for Excellence and Synergetic Innovation Center of Quantum Information and Quantum Physics, University of Science and Technology of China, Hefei, Anhui 230026, People's Repubilic of China}

\author{Xiaomei Liu}
\affiliation{Hefei National Laboratory for Physical Sciences at Microscale and Department of Modern Physics, University of Science and Technology of China, Hefei, Anhui 230036, People's Republic of China}

\author{Nanyang Xu}
\email{nyxu@ustc.edu.cn}
\author{Jiangfeng Du}
\email{djf@ustc.edu.cn}
\affiliation{Hefei National Laboratory for Physical Sciences at Microscale and Department of Modern Physics, University of Science and Technology of China, Hefei, Anhui 230036, People's Republic of China}
\affiliation{CAS Center for Excellence and Synergetic Innovation Center of Quantum Information and Quantum Physics, University of Science and Technology of China, Hefei, Anhui 230026, People's Repubilic of China}

\begin{abstract}

Machines are possible to have some artificial intelligence like human beings owing to particular algorithms or software.  Such machines could learn knowledge from what people taught them and do works according to the knowledge. In practical learning cases, the data is often extremely complicated and large, thus classical learning machines often need huge computational resources. Quantum machine learning algorithm, on the other hand, could be exponentially faster than classical machines using quantum parallelism. Here, we demonstrate a quantum machine learning algorithm on a four-qubit NMR test bench to solve an optical character recognition problem, also known as the handwriting recognition. The quantum machine learns standard character fonts and then recognize handwritten characters from a set with two candidates. To our best knowledge, this is the first artificial intelligence realized on a quantum processor. Due to the widespreading importance of artificial intelligence and its tremendous consuming of computational resources, quantum speedup would be extremely attractive against the challenges from the Big Data.

\end{abstract}
\maketitle


As an approach to artificial intelligence (AI), machine learning is computer software that have the ability to learn from data for specific tasks when no direct solutions are available \cite{intro_ml}. Machine learning is widely used in the fields such as data mining, where the amount of data grows rapidly and so does the required computational resources. There are two major types of machine learning which are supervised and unsupervised learning, differing from whether the training data are labeled in advance. Among the supervised machines, support vector machine (SVM) is a popular learning machine which learns from training data and classifies vectors in a feature space into one of two subgroups \cite{csvm}. The machine first converts the training data to feature space and find the optimal hyperplane to divide the two subgroups. Then the machine could use the hyperplane to classify a new unknown instance which subgroup it belongs. The computational complexity of SVM in time is proportional to \emph{O}(poly\emph{(NM)}), where \emph{N} is the number of dimensions of the feature space and \emph{M} is the number of the training samples. It is well-known that quantum algorithms presented in many fields provide impressive speedup over their classical counterparts \cite{Shor_algorithm,Grover_search,qcqi,rv_qc}. In the area of AI, Rebentrost \emph{et.\ al.} recently reported a quantum algorithm for SVM (QSVM)\cite{qsvm} based on recently developed techniques\cite{qpca,qml,qlinear}, which can achieve \emph{O}(log\emph{(NM)}) of performance in both training and classifying process.  The exponential speedup is achieved by utilizing a quantum circuit to calculate the inner products of the vectors in parallel\cite{qml} and convert the SVM training to an approximate least-square problem which is subsequently solved it by the quantum matrix inversion algorithm\cite{qlinear,qfit}.

Here we report an experimental implementation of the QSVM algorithm. To be more comprehensive, we apply the algorithm in a popular problem, \emph{i.e.}, a Optical Character Recognition (OCR) problem \cite{ocr}, to classify handwritten characters in a candidate set. In the state-of-art quantum computing technologies, only a limited number of qubits can be well controlled and managed. While on the other side, a realistic OCR problem in the QSVM may need tens of qubits, which can not be satisfied today. Thus here we restrict the problem to the minimal case where only two characters ( `6' and `9') are in the candidate set and only two features (horizontal and vertical ratios, which are defined in the following) are considered in the problem. This allows us to demonstrate this quantum artificial intelligence algorithm on a 4-qubit nuclear spin quantum processor at room-temperature.

\begin{figure}
\includegraphics[width=0.8 \columnwidth]{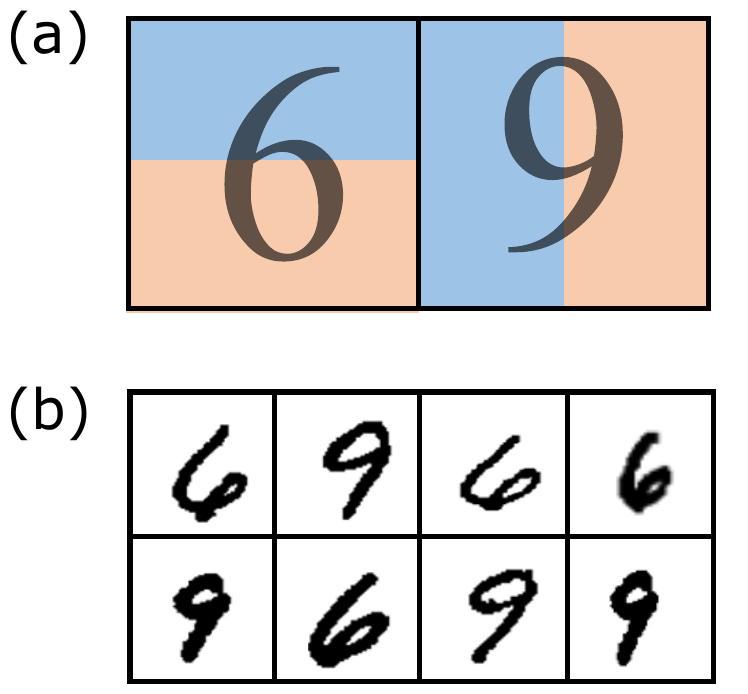}
\caption{(Color online) Samples of optical character recognition for SVM. (a) The standard font (Times New Roman) image for the two character `6' and `9' for training SVM. Features in OCR is the vertical and horizontal ratios which are calculated by the number of pixels divided by the middle line as depicted. (b) The handwritten samples for test of the SVM.}\label{fig:OCR_problem}
\end{figure}

A usual OCR process often contains four stages \cite{ocr}: (i) preprocessing the image, (ii) dividing the sentence image into character images, (iii) extracting features from the image, and (iv) classifying the character image. Here the machine is only designed to recognize a single character image, thus the second stage for segmentation is not necessary. The quantum SVM here works as follows: first train the machine by the two printed image with standard font of the characters as in Fig.\ \ref{fig:OCR_problem}a, and then the handwritten images of character `6' or `9' (as in Fig.\ \ref{fig:OCR_problem}b) are provided and the machine recognize (classify) each to determine which character group it belongs. The vertical (horizontal) ratio is calculated from the pixels in the left (upper) half over the right (lower) half (as in Fig.\ \ref{fig:OCR_problem}a). It is worth noting that, the machine actually works with a vector formed by the features of each image, and before feeding to the machine every image should be preprocessed, such as resizing the pixels. 
Furthermore we applied a linear conversion on the feature vectors to match our machine.
After these preprocessing, the two printed image with standard font can be represented by ${\vec x_1} = (0.9872,0.1595)$ for character "6" and ${\vec x_2} = (0.3544,0.9351)$ for character "9".

Before diving into the experimental details, we give a brief introduction to the SVM algorithm. The task for a SVM algorithm on classical computers is to classify a vector $\vec{x}_0$ into one of two classes which is denoted by a variable $y_i=\pm1$  for each class. In the training process where $M$ training samples in the form $\{(\vec{x}_i,y_i)\}$ are provided, the machine finds a maximum-margin hyperplane with normal vector $\vec{w}$  which separates the samples by their classes. The margin in the vector space is bounded by two parallel hyperplanes with maximal possible distance $2/|\vec{w}|$ and no points inside. Mathematically, we construct the hyperplane with $\vec{w}\cdot\vec{x}+b=0$ so that $\vec{w}\cdot\vec{x}_i+b\geqslant1$ for $\vec{x}_i$ in the $+1$ class and $\vec{w}\cdot\vec{x}_i+b\leqslant-1$ for $\vec{x}_i$ in the $-1$ class where $b/|\vec{w}|$ is the offset of the hyperplane. Thus in this formation, to find the optimal hyperplane is to minimize $|\vec{w}|^2/2$ subject to the constraints $y_i(\vec{w}\cdot\vec{x}_i + b)\geqslant1$ for all $i$.  This dual formulation is to maximize the function $L(\vec{\alpha})=\sum_{i=1}^{M}{y_i\alpha_i} - \frac{1}{2}\sum_{i,j=1}^{M}{\alpha_iK_{i,j}\alpha_j}$ over the Karush-Kuhn-Tucker multipliers $\vec{\alpha}$ subject to the constraints $\sum_{i=1}^{M}{\alpha_i=0} $ and $ y_i\alpha_i\geqslant0$ \cite{dual}.

The hyperplane parameters $\vec{w}$ and $b$ could be recovered from the least-squares approximation of SVM \cite{LS_SVM}:
\begin{equation}\label{eq:Linear_equation}
\tilde F \begin{pmatrix} b  \\ \vec{\alpha}  \end{pmatrix}\equiv\\
\begin{pmatrix} 0 & \vec{1}^T  \\ \vec{1}  & K+\gamma^{-1}\mathbf{I}_M\end{pmatrix}\\
\begin{pmatrix} b  \\ \vec{\alpha}  \end{pmatrix}=\\
\begin{pmatrix} 0  \\ \vec{y}  \end{pmatrix},\\
\end{equation}
where $K$ is the kernel matrix with $K_{i,j}=k(\vec{x}_i,\vec{x}_j)=\vec{x}_i\cdot\vec{x}_j$, $y=(y_1,\cdots,y_M)^T$, $\vec{1}=(1,\cdots,1)^T$ and $\mathbf{I}_M$ is the $M \times M$ identity matrix. Here $\gamma$ determines the relative weight of the training error and the SVM objective. Solving this linear equation leads to the value of offset $b$ and the normal vector $\vec{w}=\sum_{i=1}^{M}\alpha_i\vec{x}_i$. Then the classification process for the handwritten character represented by $\vec{x}_0$ is $y(\vec{x}_0)=\mathbf{sign}(\sum_{i=1}^{M}\alpha_ik(\vec{x}_i,\vec{x}_0)+b)$, which also have a matrix form:
\begin{equation}\label{eq_classification}
y(\vec{x}_0) = {\bf{sign}}\left [\left( {\begin{array}{*{20}{c}}
1&{{{\vec x}_0}}& \cdots &{{{\vec x}_0}}
\end{array}} \right)\left( {\begin{array}{*{20}{c}}
b\\
{{\alpha _1}{{\vec x}_1}}\\
 \vdots \\
{{\alpha _M}{{\vec x}_M}}
\end{array}} \right)\right].
\end{equation}
In the OCR problem, we will define the character "6" as "positive class" , \emph{i.e.} $y(\vec x_1) =  + 1$, and the character "9" as "negative class", \emph{i.e.} $y(\vec x_2) =  - 1$.

The process of Least-squares SVM could be summarized as follows: (i) calculate the kernel matrix $K$, (ii) solving the LS-equation to obtain the hyperplane parameters, and (iii) the classification result will be produced from Eq.\ (\ref{eq_classification}), which is the inner product of a query-vector and a training-data vector.

Stepping into the quantum SVM, we need to assume that we have the training-data oracles that could prepare the state $|\vec{x}_i\rangle=1/|\vec{x}_i|\sum_{j=1}^{N}(\vec{x}_i)_j|j\rangle$, which is the quantum counterpart of the training data ${\vec x_i}$. Starting from the initial state $1/\sqrt M \sum_{i = 1}^M {\left| i \right\rangle }$, the training-data oracles are used to prepare the state $\left| \chi  \right\rangle  = 1/\sqrt {{N_\chi }} \sum\nolimits_{i = 1}^M {\left| {{{\vec x}_i}} \right|\left| i \right\rangle \left| {{{\vec x}_i}} \right\rangle }$, with ${N_\chi } = \sum\nolimits_{i = 1}^M {{{\left| {{{\vec x}_i}} \right|}^2}}$. After discarding the training set register, the quantum density matrix will reduce to the kernel matrix up to a constant factor, \emph{i.e.}, $\mathbf{tr}K$ .

In the second step, the quantum algorithm for solving linear equation have been proposed \cite{qlinear} and experimentally realized \cite{LEexp_optics,LEexp_NMR}, with an exponentially speedup. Using the same method, the hyperplane parameters are determined by ${\left( {b,{{\vec \alpha }^T}} \right)^T} = {\tilde F^{ - 1}}{\left( {0,{{\vec y}^T}} \right)^T}$, where the vectors here represent quantum states.

\begin{figure}
\includegraphics[width= 1 \columnwidth]{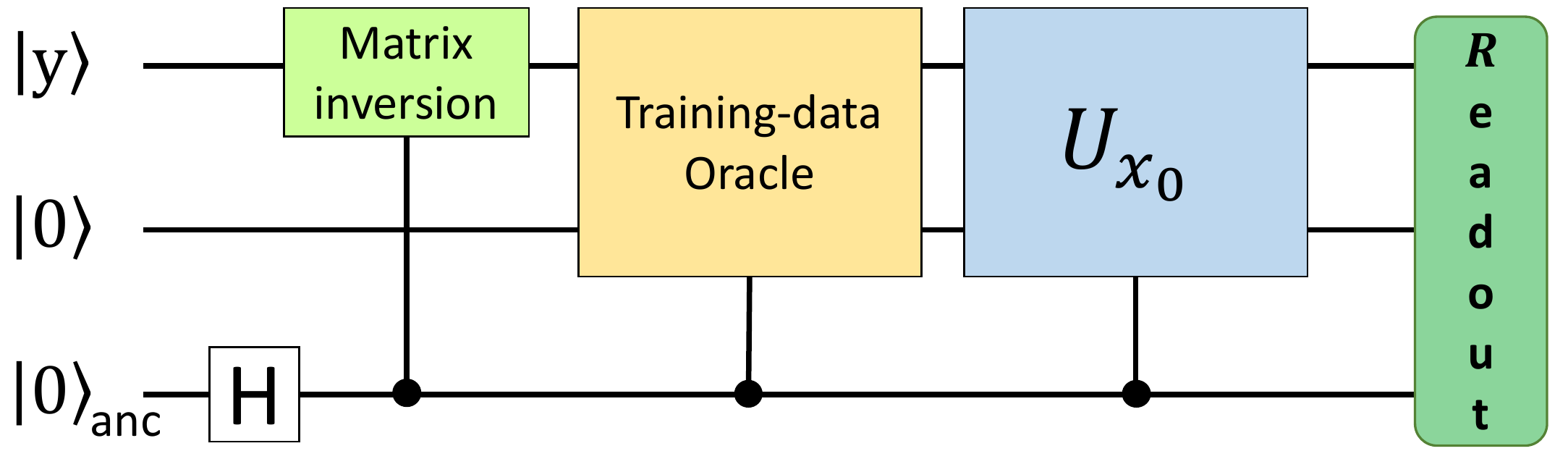}
\caption{(Color online) The schematic diagram of the quantum SVM. An ancillary qubit is added here to readout the classification result. The auxiliary registers for matrix inversion are not shown here. }\label{Fig:Classification}
\end{figure}

\begin{figure*}
\includegraphics[width= 1 \columnwidth]{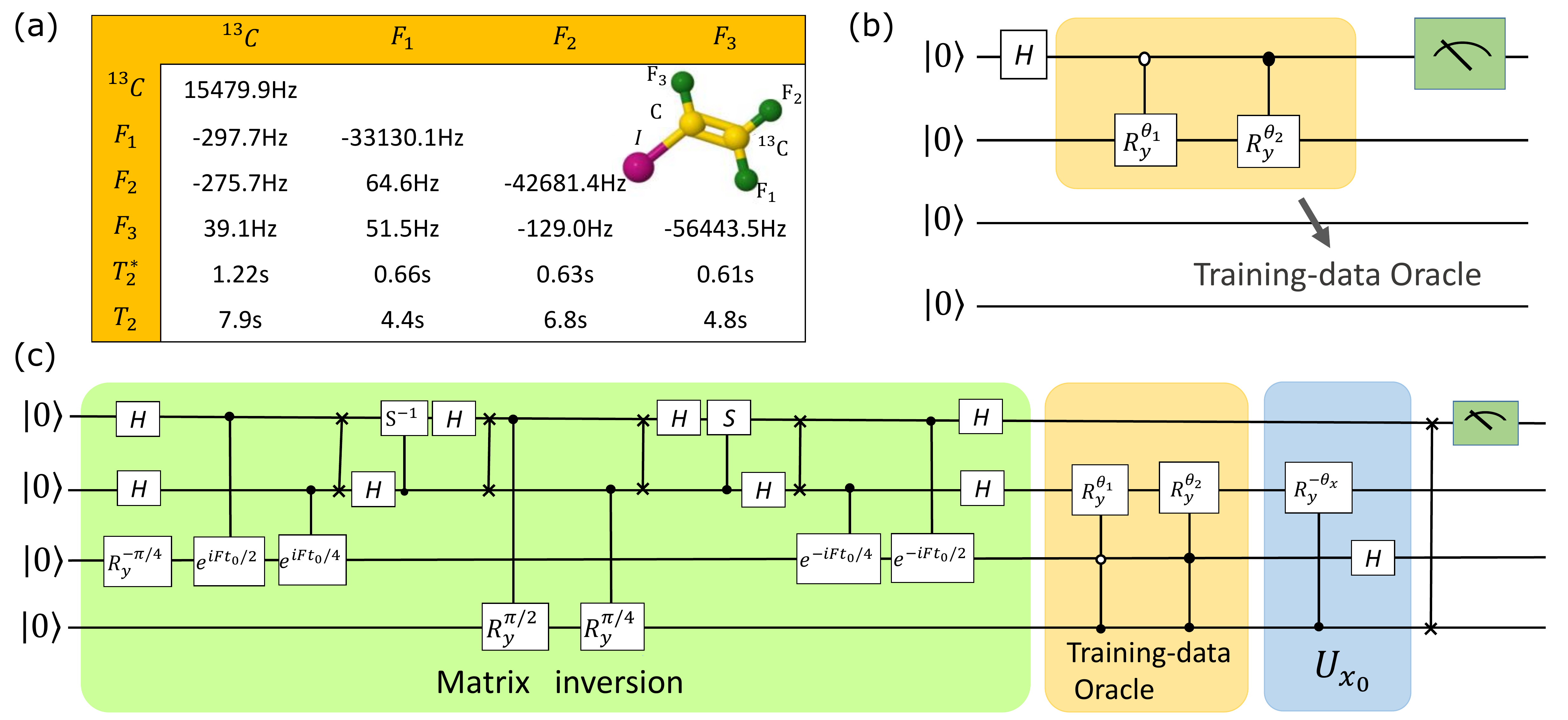}
\caption{(Color online)(a) Properties of the $^{13}$C-iodotrifluroethylene. The chemical shifts $\nu_i$ and scalar coupling constants ($J_{jk}$) are on the lower diagonal in the table, respectively. The chemical shifts are given with respect to reference frequencies of 100.62 MHz (Carbon) and 376.48 MHz (Fluorines). (b) The quantum circuit for building the kernel matrix $K$. After discarding the training-data register (the second qubit), the desired kernel matrix $K$ is obtained as the quantum density matrix of the first qubit. (c) The quantum circuit for classification. Here $H$ and $S$ are the Hadamard and phase gate, respectively.
}\label{Fig:circuits}
\end{figure*}

The classification results in Eq.\ (\ref{eq_classification}) could be reproduced by the overlap of two quantum states : $y(\vec x) = {\bf{sign}}(\left\langle {{{\tilde x}_0}\left| {\tilde u} \right.} \right\rangle )$, with the training-data state $|\tilde u\rangle  = \frac{1}{{\sqrt {{N_{\tilde u}}} }}(b|0\rangle |0\rangle  + \sum_{k = 1}^M {\textrm{abs}(\vec x_k)} {{\alpha _k}}|k\rangle |{\vec x_k}\rangle )$ and the query-state $|\tilde{x_0}\rangle=\frac{1}{\sqrt{N_{\tilde{x_0}}}}(|0\rangle|0\rangle + \sum_{k=1}^{M}{\textrm{abs}(\vec x_0)}|k\rangle|\vec{x_0}\rangle)$. Here the training-data state $|\tilde{u}\rangle$ could be easily obtained by calling the training-data oracle on ${\left( {b,{{\vec \alpha }^T}} \right)^T}$. By applying a inverse operation ${U_{{x_0}}} = \left| {00} \right\rangle \left\langle {{{\tilde x}_0}} \right|$, the expansion coefficients $\left\langle {00} \right|{U_{{x_0}}}\left| {\tilde u} \right\rangle  = \left\langle {{{\tilde x}_0}} \right|\left| {\tilde u} \right\rangle $ will produce the classification result $y(\vec x)$ \cite{note1}. A schematic diagram of this part is shown in Fig.\ \ref{Fig:Classification}. Note that the unitary operations are conditional operations here, controlled by an ancillary qubit. Hence the final state will be $\left| \psi  \right\rangle  = \left| \phi  \right\rangle {\left| 1 \right\rangle _A} + \left| {00} \right\rangle {\left| 0 \right\rangle _A}$, where $\left| \phi  \right\rangle  = {U_{{x_0}}}\left| {\tilde u} \right\rangle $ and the subscript "A" indicates the state of ancillary qubit. By measuring the expectation value of coherent term $O \equiv \left| {00} \right\rangle \left\langle {00} \right| \otimes {\left( {\left| 0 \right\rangle \left\langle 1 \right|} \right)_A}$, the classification result will be revealed by:
\begin{equation}\label{eq_classification_result}
{y({{\vec x}_0}) = {\bf{sign}}(\left\langle {00\left| \phi  \right.} \right\rangle ) = {\bf{sign}}\left( {\langle \psi {\rm{|}}O\left| \psi  \right\rangle } \right).}
\end{equation}
If the expectation value is greater than 0, the classification result will be positive ($y(\vec x_0) =  + 1$), otherwise it will be negative ($y(\vec x_0) =  - 1$).

Now we will turn to the experimental implementation of the quantum algorithm for OCR. We employ the widely-utilized nuclear magnetic resonance (NMR) technique \cite{nmr1,nmr2,molsim,cdsim,compress} to realize the 4-qubit quantum processor. The experiments are carried out on a Bruker AV-400 spectrometer at 306 K. The sample used is the $^{13}$C-iodotrifluroethylene(${{\rm{C}}_2}{\rm{F}_3}\rm{I}$) dissolved in d-chloroform, which consists of a $^{13}$C nuclear spin and three $^{19}$F nuclear spins. The natural Hamiltonian of this weakly-coupled spin system in the rotating frame is
\begin{equation}\label{HNMR}
{H_{{\rm{NMR}}}} = \sum\limits_{j = 1}^4 {\pi {\nu _j}} \sigma _z^j + \sum\limits_{1\leq j < k \leq  4} {\frac{\pi }{2}} J_{jk}\sigma _z^j\sigma _z^k,
\end{equation}
with the parameters shown in Fig.\ \ref{Fig:circuits}a.  The deviation density matrix of thermal equilibrium state is ${\rho _{\rm eq}} = \sum_{i = 1}^{4} {{\gamma _i}\sigma_z^i} $, where $\gamma_i$ represents the gyromagnetic ratio of each nuclear spin.

The experimental procedure consists of three parts: (i) pseudopure state (PPS) preparation \cite{pps1}, (ii) building the kernel matrix K, and (iii) machine learning and classification. Starting from the thermal state ${\rho _{{\rm{eq}}}}$, the line-selective method is employed to prepare the PPS \cite{PPS_preparation}: ${\rho _i} = \frac{{1 - \varepsilon }}{{16}}\mathbf{I}_{16} + \varepsilon \left| {0000} \right\rangle \left\langle {0000} \right|$. Here $\mathbf{I}_{16}$ represents the $16 \times 16$ identity operator and $\varepsilon  \approx {10^{ - 5}}$ is the polarization.

The quantum circuit of the second part is shown in Fig.\ \ref{Fig:circuits}. Here the first qubit ($^{13}$C ) serves as the probe qubit, and the second qubit (training set register) is encoded with information of the two standard sample of the characters "6" and "9". The training-data oracle is realized by using two controlled rotations here.
After that, the state of the probe qubit will represent the kernel matrix $K$, up to a constant factor $\mathbf{tr}K$. In the experiment, the density matrix $K/\mathbf{tr}K$ is measured as $\left( {\begin{array}{*{20}{c}}
{0.5065}&{0.2425}\\
{0.2425}&{0.4935}
\end{array}} \right)$ by utilizing the quantum state tomography technique \cite{tomo}.
%

The third part of the experiment is classification. The training set are encoded into the system register (the third qubit). In this implementation, we will make the non-offset reduction, \emph{i.e.}, $b=0$. Then the linear equation (Eq.\ (\ref{eq:Linear_equation}))will reduce to $F\vec \alpha  \equiv \left( {K + {\gamma ^{ - 1}}\mathbf{I}_2} \right)\vec \alpha  = \vec y$, which is $2 \times 2$ dimensional in this case. The user-specified weight is set as $\gamma=2$. Here by using a rotation along $y$ axis, \emph{i.e.} $R_y^{-\pi/4} $, the system register are prepared into $\left| y \right\rangle {\rm{ = }}\left( {\begin{array}{*{20}{c}}
{\rm{1}}\\
{{\rm{ - 1}}}
\end{array}} \right){\rm{/}}\sqrt {\rm{2}} ,$. Then the phase estimation algorithm is used to realize the matrix inversion \cite{qlinear,qfit}, which generate the state  ${F^{ - 1}}\left| y \right\rangle $.  In this stage, the weight of support vectors $\alpha_i$ have been solved and stored into the system register, when the forth qubit is on state $\left| 1 \right\rangle $. By calling the training-data oracle again, the weights are encoded to the coefficient of related support vectors $\left| {{x_j}} \right\rangle $, leading to the preparation of training-data state $|\tilde u\rangle $. Then the inverse operation $U_{x_0}$, which relates to the query state $x_0$, is applied. Note that these operations after matrix inversion ${F^{ - 1}} $ should be conditional rotation, as shown in Fig.\ \ref{Fig:Classification}. In experiment, we pack the circuit for matrix inversion into one shaped pulse optimized by the gradient ascent pulse engineering (GRAPE) method \cite{GRAPE_method}, with the length of each pulse being 20 ms and the number of segments being 1000. The rest parts of the circuit is packed into another GRAPE pulse, with the length being 25 ms and 2000 segments. All the pulses have theoretical fidelities over $99\%$ and are designed to be robust against the inhomogeneity of radio-frequency pulses.

\begin{figure}
\includegraphics[width= 1 \columnwidth]{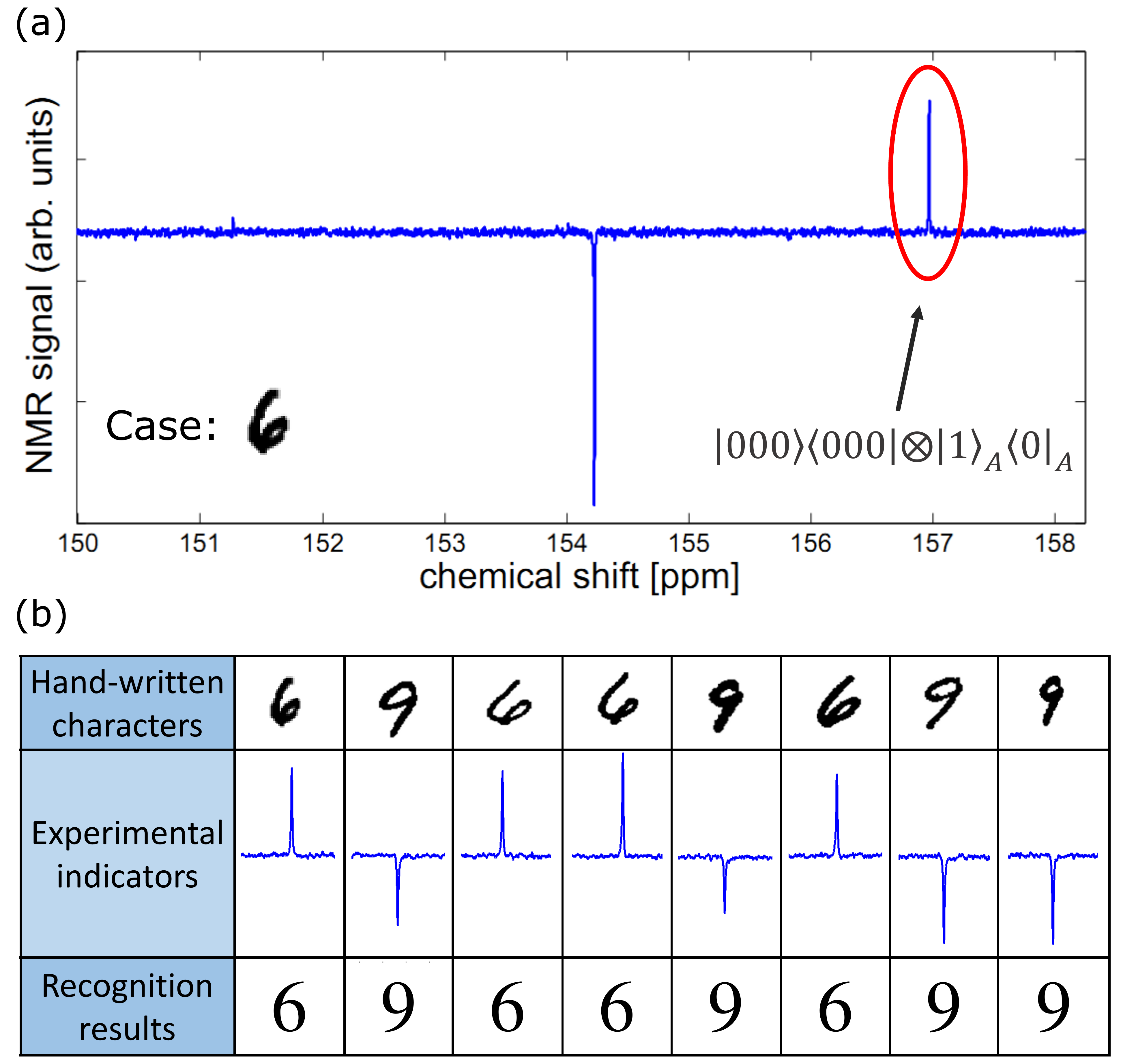}
\caption{(Color online)(a) The recognition results given by the quantum OCR, indicated by the orientation of the labeled peak in $^{13}$C spectrum. The upward peak in this spectrum presents a positive result, which classifies the incoming hand-written character as "6". (b) The recognition results corresponding to the hand-written characters in
Fig.\ \ref{fig:OCR_problem}b. The three rows represent the hand-written characters, the experimental indicators and the recognition results subsequently.
}\label{Fig:Results}
\end{figure}

Finally, the classification result is read out through the expectation value of the coherent term of the ancillary qubit $\left| {000} \right\rangle \left\langle {{\rm{000}}} \right| \otimes {\left| 1 \right\rangle _A}{\left\langle {\rm{0}} \right|_A}$. The expression is slightly different from Eq.\ (\ref{eq_classification_result}), since the auxiliary registers for matrix inversion are not shown there. In the experiment, the information of the ancillary qubit is transferred to the $^{13}$C spin by a \rm{SWAP} gate, and then read through the $^{13}$C spectrum \cite{C2F3I}. If the corresponding peak in $^{13}$C spectrum is upward (\emph{e.g.} Fig.\ \ref{Fig:Results}), the classification result will be positive, which means the character recognition results is "6". On the other hand, when the peak is downward, the classification result will be "9". In the experiment, we perform 8 different recognition tasks, corresponding to the 8 hand-written characters in Fig.\ \ref{fig:OCR_problem}. The results are shown in Fig.\ \ref{Fig:Results}b, which accord with human observation.

%

%
To conclude, in this letter we demonstrate the first artificial intelligence algorithm implemented quantum-mechanically on a four-qubit quantum processor, \emph{i.e.}, a quantum learning machine. As an example, we utilized the quantum support vector machine to solve a minimal optical character recognition problem. In the experiment, the quantum machine are trained with standard font of the characters "6" and "9", and then used to classify new-coming hand-written characters. The successful classification shows the ability of our quantum machine to learn and work like an intelligent human. This quantum machine could be extended to large-scale dataset using resources only logarithmic in feature size and the number of training data. This work paves the way to a bright future where the Big Data is processed efficiently in a parallel way provided by quantum mechanics.

%


%
%


This work was supported by the National Key Basic Research Program of China (Grant No.\ 2013CB921800),
the National Natural Science Foundation of China (Grant No.\ 11227901, No.\ 91021005 and No.\ 61376128), the Strategic Priority Research Program (B) of the CAS (Grant No. XDB01030400) and the Fundamental Research Funds for the Central Universities.


\end{document}